\documentclass[a4paper,10pt]{article}

\usepackage{a4wide}
\usepackage{graphicx,setspace}
\usepackage{bm}
\usepackage{amsmath,textcomp}
\usepackage{amssymb}
\usepackage{subfig}
\usepackage{xcolor}

\begin{document}
\vspace*{-3cm}
\begin{flushleft}
{\Large
\textbf{The Detection and effect of social events on Wikipedia data-set for studying human preferences}
}
\\
Julien Assuied$^{1}$,
Y\'{e}rali Gandica$^{2,\ast}$,
\\
\singlespacing
\footnotesize {
{1} CY-Tech  Cergy Paris Université.\\
{2} CY Cergy Paris Université, CNRS, Laboratoire De Physique Théorique et Modélisation, F-95000 Cergy, France.
}
\end{flushleft}

\singlespacing
\begin{abstract} 
Several studies have used Wikipedia (WP) data-set to analyse worldwide human preferences by languages. However, those studies could suffer from bias related to exceptional social circumstances. Any massive event promoting the { exceptional} edition of WP can be defined as a source of bias. In this article, we follow a procedure for detecting outliers. Our study is based on  $12$ languages and $13$ different categories. Our methodology { defines} a parameter, which is language-depending instead of being externally fixed. We also study the presence of human cyclic behaviour to  {evaluate} apparent outliers. After our analysis, we found that the outliers in our data set do not significantly affect using the { whole} Wikipedia-data set as a digital footprint to analyse worldwide human preferences.
\end{abstract}

\singlespacing
\section*{Introduction}
Editing the worldwide encyclopedia Wikipedia (WP) has several important implications. First of all, it is nowadays an essential and primary source of knowledge, accessible to everybody with access to the Internet. Secondly, that source of information does not follow an editorial line. Thus, different perspectives and points of view on the same subject can be read in an article if it is well structured and democratised. The third point, which is the one that concerns us in this article, is that the information is a footprint of the community's cultural context. For example, by a cross-language analysis of over 40 languages, Miquel-Ribé et al. identified that at least a quarter of a WP page's content is generally dedicated to the specific cultural context \cite{fphy}. That study confirms the inter-cultural nature of Wikipedia as a digital repository of human cultural context. In the same order of ideas, Mehler et al. have shown that information on a given topic is strongly language-dependent \cite{feduc}. { Even the definition for the quality of an article depends on the WP language as reported by Jemielniak et al.\cite{Jemielniak&Wilamowski}}. The presence of those important content imbalances among languages {suggests} the success of WP as a good proxy for analysing cultural context or group preferences.

In that respect, several studies regarding worldwide human attributes have been performed using the Wikipedia data-set. For example, Samolienko et al. studied the similarity of interests between cultural communities and cross-lingual interconnections \cite{epj}. In a previous work \cite{SNAMS}, we studied worldwide human preferences showing, for example, which languages have uniform preferences among subjects, in contrast with the ones showing pronounced specific inclinations.

All those studies could suffer from some bias related to exceptional social circumstances, for example, special events, accidents; any conjuncture that can alter the regular pattern of editing WP based on { self-choice}. When using Wikipedia data-set for analysing human preferences, any massive event promoting the { extraordinary} edition of WP can be defined as a source of bias. In Data Science, the term for such bias is outliers. In general, outliers are data points that deviate markedly from the trend in the sample \cite{book}. Those outliers could result from errors in the data, for example, errors during the encoding or even during the experimental set-up. In that case, such points could be deleted from the data. However, outliers could also be a consequence of a considerable variance, in which case, it could point out some interesting phenomenon.  

An even more difficult situation is when the whole data-set presents bias due to massive actions, like special events or natural disasters. Then, if the scientific question is related to legitimate human preferences, those events could cause serious bias and cannot be detected as singular points due to their massive nature. { In this work}, we propose a methodology to deal with such a situation. Our methodology allows us to answer the question: Are the preferences between individuals sharing the same language, which is a trace of the collective identity of the entire group, the same whenever the effect of extraordinary events is removed?

The analysis has two parts. The first part deals with identifying outliers from monthly data points. Our methodology does not depend on any external threshold; it depends intrinsically on each language. The second part aims to detect periodicity, and we use the Fourier transform for that endeavour.

Our study is based on the first ten years of Wikipedia data-set to avoid the artificial process of editing WP caused by the recommendation system that was applied by the Wikimedia Foundation to fill that gap caused by inner cultural preferences \cite{gap}. Our analysis considers twelve languages: the ones written in Spanish (ES-WP), French (FR-WP), Portuguese (PT-WP), Italian (IT-WP), Hungarian (HU-WP), German (DE-WP), Russian (RU-WP), Arabic (AR-WP), Japanese (JA-WP), Chinese (ZH-WP) and Vietnamese (VI-WP). {Our selection takes into account the interplay between a worldwide view and the WP sizes.} 

\section*{The data-set}
The activity of the first ten years of WP editing activity WP was downloaded. The starting points depend on each WP language, and they are in the range from 11/10/2001 to 28/03/2010. Bot activities were removed by detecting users whose name contains the word bot in either combination of both uppercase and lowercase. 

\section*{The data-set}
The activity of the first ten years of WP editing activity WP was downloaded. The starting points depend on each WP language, and they are in the range from 11/10/2001 to 28/03/2010. Bot activities were removed by detecting users whose name contains the word bot in either combination of both uppercase and lowercase. Table \ref{tab:twitternets} shows the number of pages, users and edits, and the starting date for each WP data set used in this study.

\begin{table}[h!]
\centering
\begin{tabular}{|c|c|r|r|r|r|}
\hline
 WP & \# pages & \# users & \#  edits & \# starting date\\
 \hline
 $ES$ & 1144177 & 2682095 & 47728243  &  24/09/2001 - 17:01:24  \\
 $FR$ & 2936383 &  203038 & 58325545 &  13/10/2001 - 09:59:23  \\
 $PT$ & 894521 &  1475236 & 19937771 &  17/06/2001 - 17:13:19 \\
 $IT$ & 1084333 &  97161 & 22200807  &  14/09/2001 - 10:19:28 \\
 $HU$ & 248808 &  148068  & 5758998  & 09/07/2003 - 04:41:24 \\
 $DE$ & 1111265 &  357561  & 39689676  &   09/09/2001 - 03:34:41\\
 $RU$ & 1134752 &  76066  & 14199590 & 13/11/2002 - 18:48:05 \\
 $AR$ & 624118 &  23641  & 7674946 &  11/07/2003 - 00:23:02  \\
 $JA$ & 690795 &  126657  & 24584471 &  10/09/2002 - 19:25:48  \\
 $ZH$ & 495855 &  76613  & 3657770 &  30/10/2002 - 17:19:19  \\
 $VI$ & 238859 &  12262  & 12618296 &  16/11/2002 - 14:54:24  \\
 \hline
\end{tabular}
\caption{Data-set.}
\label{tab:twitternets}
\end{table}

In order to have a broad spectrum of preferences among individuals sharing the same language, we used the categories already classified in the main branch of the tree structure, defined by the Wikimedia Foundation itself, and found in \cite{tree}. To avoid a substantial overlap between categories, we left out of this study the categories: 
\begin{itemize}
 \item Culture,
 \item Humanities,
 \item Law,
 \item Life,
 \item Matter,
 \item People,
 \item Reference Works, 
 \item Science and Technology, 
 \item Society, 
 \item Universe, 
 \item World. 
\end{itemize}
Thus, our study is based on the $13$ following categories: Arts, Sports, Right, Events, Philosophy, Geography, History, Games, Mathematics, Nature, Politics, Religion, and Health. We used the Petscan API \cite{API} in order to download the names of all the pages within each category in the 12 languages under study. Notice that different languages have different pages in the corresponding category.

\section*{Trending}

\begin{figure*}[h]
\centering
\includegraphics[width=\linewidth]{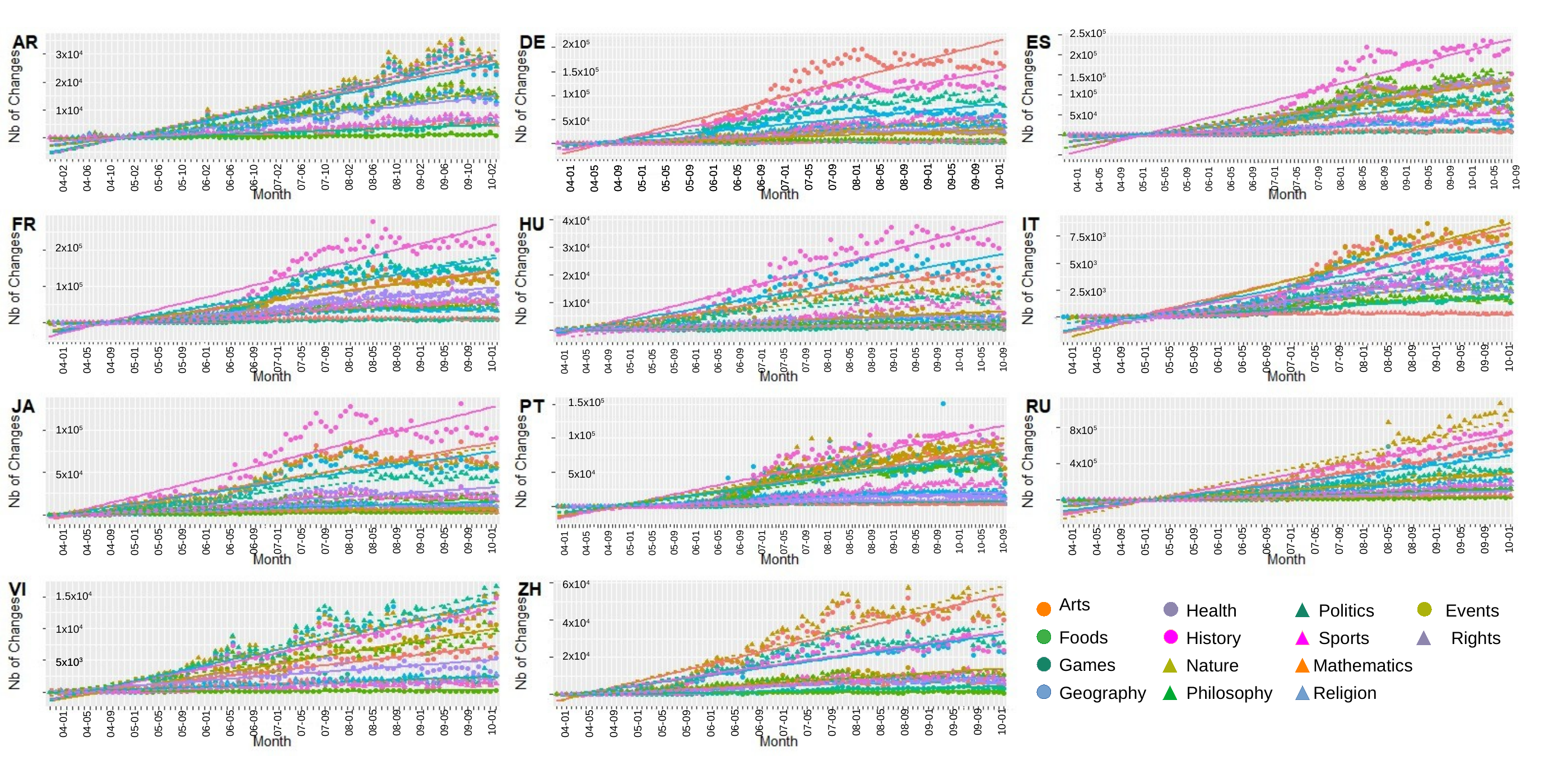}
\caption{\label{editions} \textbf{Trending:} Number of editions per months according to the categories, as represented by the legend. A trend line for each data set has been added. Each plot refers a different WP language.  Three different behaviours have been identified, see text.}
\end{figure*}

Let us start visualising the data. In figure \ref{editions}, we show, for each language, the number of editions per month according to the categories.  After playing with the axes, we realised that there are no power-law or exponential trending for the whole period in any curve. We got that points mostly move around linear trends in all the curves. For that reason, we have added a trend obtained from the linear regression to better visualise the behaviour of the data for each category. However, let us notice that some curves present a cyclic behaviour around those coarse-grained linear trends.

We can identify three different behaviours. Some data-set mostly follow their trends, as in most WP-RU categories. In some other cases, the points oscillate around the trends, as in WP-AR and WP-VI. Finally, some curves get far from the trend, such as History in WP-FR and WP-JA and Arts in WP-DE; {however, they cross the linear trend several times.}

We will analyse the last two cases separately. Outliers points will be located and eliminated from the data set where the curves are considerably separated from the trends. On the other hand, the case of oscillatory data will be inspected by employing a Fourier analysis to detect their periodicity.

\section*{Finding Outliers}
We need to spot outliers for the data points far from the trend. For that endeavour, we first must define them. { In order to decrease the bias caused by the researcher, we aim at fixing the threshold directly from the data rather than being arbitrarily settled. Then, in the same spirit that two points are necessary to define a straight line, here we use two data sets to fix a language-depending threshold. The necessary condition on the two data sets is that they should suffer from the same bias, which in our work are extraordinary events.}

{\ The two data sets that we use for setting the threshold are the monthly time series for the number of edits (shown in figure 1) and the monthly time series for the number of users. We hypothesise that during extraordinary massive events, extra editions are performed by regular editors, but also new editors enter information to WP.} 

\begin{figure*}[h]
\centering
\includegraphics[width=\linewidth]{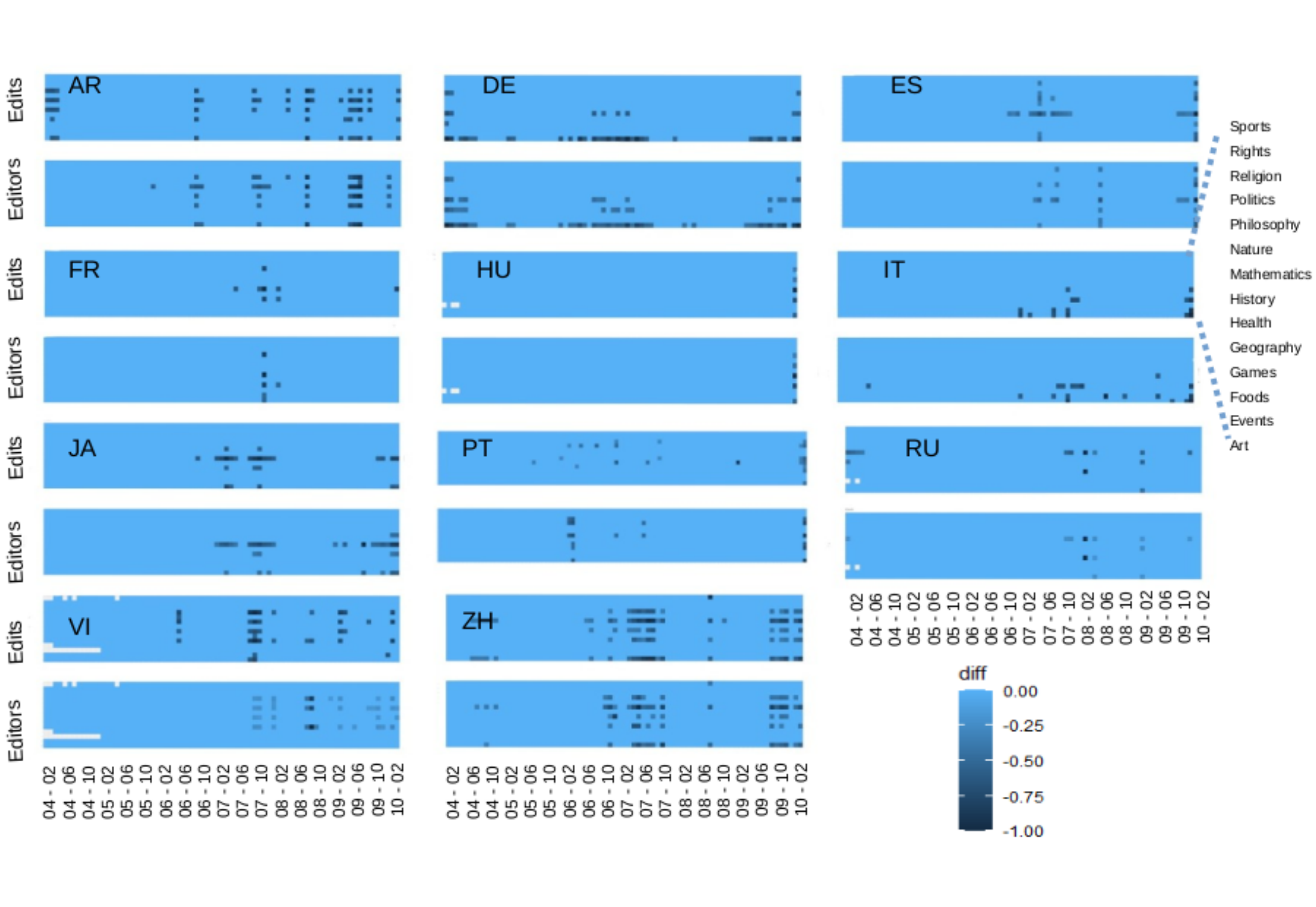}
\caption{\label{heatmap} \textbf{Finding Outliers:} Heat-maps showing the difference between the data points and trends for edits and editors, normalised in terms of the highest value of each graph. Each horizontal line is one category as signalled in the plot for Edits-IT. The heat-map legend is shown at the bottom right.}
\end{figure*}

{Then, the difference between the monthly points and their trends values are calculated for each category. We normalised the values of these differences in terms of the highest value of each graph. In this way, we generated, for each language, one plot for editors and one plot for editions, which are the heat-maps shown in figure \ref{heatmap}.}


{ Next, for each language, we compared the two heat-maps: editors and editions. Finally, we took as the threshold the minimum value where those differences appear on both plots. All the data points whose difference with the trend is greater than that threshold value are defined as outliers. Let us notice that in figure \ref{heatmap}, each WP language has two plots. The one above (heat-map for edits) is the one which should be compared to figure \ref{editions} (number of editions per month). }

{We have already identified outliers as the points that are far from the trends. However, we have seen that some curves present cyclic behaviour. Actually, human behaviour is characterised by cyclic patterns \cite{gandica2016}. In other words, periodic behaviour is intrinsic to human nature and should not be identified as extraordinary situations; some examples are national days and commemoration of great tragedies. Following this order of ideas,} if some data points were identified as outliers and result from intrinsic human cyclical patterns, we should not delete them from the data. The study for that cyclical behaviour is performed in the next section.

\section*{Cyclical Analysis}

As said before, some points distant from the trend show periodic behaviour. As cyclical patterns are an essential feature of human activity, we will not consider them outliers. However, it is important to recognise the strength of periodicity.

\begin{figure*}[h]
\centering
\includegraphics[width=\linewidth]{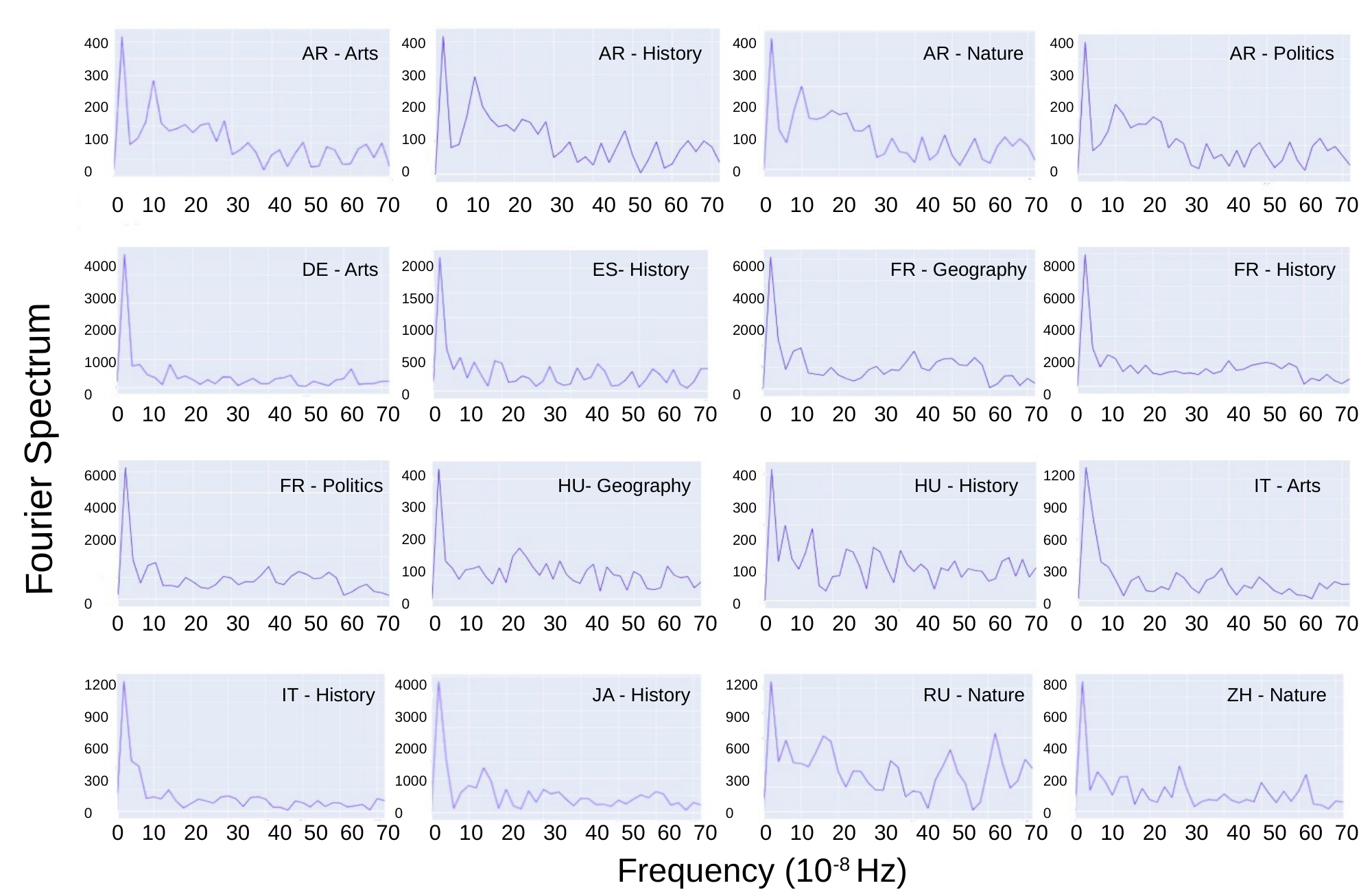}
\caption{\label{fft} \textbf{Cyclical Analysis:} Fast Fourier Transform for the temporal series where the amplitude is considerably large compared to the other categories of the same language. All the figures show a cyclic behaviour of 12 months, and only the AR-WP presents a second pick for a period of three months.}
\end{figure*}

In figure \ref{fft} we show the Fast Fourier Transform (FFT) for the temporal series where the amplitude is considerably large compared to the other categories of the same language. All the figures show a cyclic behaviour of 12 months ($ \approx 3.2 \times 10^{-8} seg^{-1}$). Only the AR-WP presents a second pick for a three-months period  ($ \approx 13 \times 10^{-8} seg^{-1}$). Notice that the input for the FFT is the original time series in seconds and not the data shown in figure \ref{editions}. We also found out daily periodicity; however, we do not report them because they are related to the expected daily human cycle, thus not interesting for the present study.

These signs of periodicity allow us to identify some apparent outliers on the heat-maps, which could have been considered as extraordinary events; however, their periodicity indicates that they are consequences of regular human periodic patterns (deeper than simple daily/weekly cycles).

\section*{Results}

After discarding cyclic behaviours as possible outliers, we identified two of the external events. { The first one in January 2008 in Italy seems to have a link with an artistic event, then affecting the categories  ``Arts" and  ``Events".} The second event that we identified was in the Portuguese Wikipedia. We observed a peak of modification concerning pages related to geography and nature in October 2007. These articles follow the big fires in the Amazonian forest in Brazil, a Portuguese-speaking country.

\begin{figure*}[h]
\centering
\includegraphics[width=\linewidth]{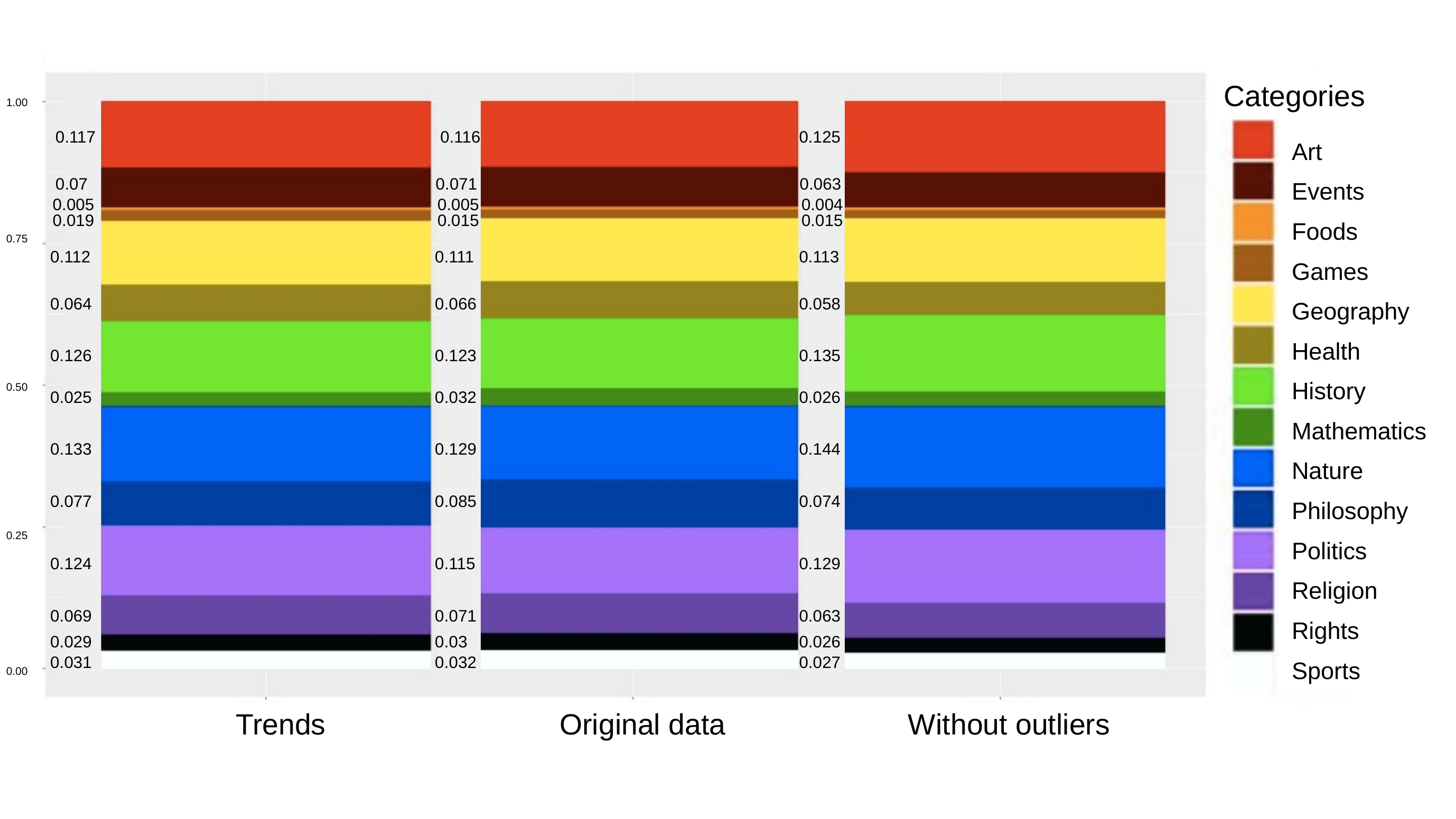}
\caption{\label{values} \textbf{Comparison:} Normalized means of the ratio of categories per month on the VI-WP. We show the values for the trends, the original values and the values without outliers.}
\end{figure*}

To better identify all the external circumstances affecting the digital footprints of Wikipedia as a social repository to study human preferences by languages, {we can see the top-edited WP pages; however, such an analysis is out of the scope of this work. On the contrary, our goal is to detect those events as outliers manifested by the same Big-Data analysis, i.e., letting the data speak.

Finally, in figure \ref{values}, we show, as one of the examples, the proportions of the categories in the total sample of the Vietnamese Wikipedia (VI). We show the proportions for the trends, the original proportions (both shown in figure \ref{editions}), and also the proportions after removing the outliers, as explained throughout this manuscript. After computing the p-values of those data, we deduced that there is no significant difference, and the same happened for all the languages. In that respect, we show that the extraordinary events on the Wikipedia data-set are non-significant when using this data-set to analyse language preferences. Our study covers the first ten years of some languages in Wikipedia as a data-set.  }    

\section*{Conclusions}

In this paper, { we first explained the importance of detecting possible bias related to extraordinary massive events when using any data set to analyse human patterns. Then, we develop a procedure to detect those possible outliers on the Wikipedia data set. Next, we analysed cyclicality to separate outliers from typical human cyclical patterns, which should not be confused with atypical behaviours, namely outliers.}

After processing the data, as explained above, we computed the proportions of categories as signs of preferences by languages and found no significant difference.    

Even though investigating the role of outliers is always a safe measure before performing any analysis in Big Data, at least in the case of the use of Wikipedia editions, during the first ten years (as we have used it here), the outliers do not have a significant effect on the results. Our primary interest was describing the preferences among the different languages in the categories, as done originally in \cite{SNAMS}.  

\section*{Acknowledge}
This work was supported by the OpLaDyn grant obtained in the 4th round of the Trans-Atlantic platform Digging into Data Challenge (2016-147 ANR OPLADYN TAP-DD2016).

\end{document}